\documentclass[%
 reprint,
 amsmath,amssymb,
 aps,
]{revtex4-1}

\usepackage{graphicx}
\usepackage{dcolumn}
\usepackage{bm}
\usepackage{braket}
\usepackage{amsmath}
\usepackage{amssymb}
\usepackage{feynmp}
\usepackage{graphics}
\usepackage{color}
\usepackage[normalem]{ulem} 

\newcommand{\ring}{\rotatebox[origin=c]{122.5}{$\lozenge$}}

\begin{document}


\title{Non-local interactions in moir\'e Hubbard systems}

\author{Nicol\'as Morales-Dur\'an$^{1}$} 
\author{Nai Chao Hu$^{1}$}
\author{Pawel Potasz$^{2}$}
\author{Allan H. MacDonald$^{1}$}
\affiliation{$^1$Department of Physics, University of Texas at Austin, Austin, Texas, 78712, USA}
\affiliation{$^2$Institute of Physics, Faculty of Physics, Astronomy and Informatics, Nicolaus Copernicus University, Grudziadzka 5, 87-100 Toru\'n, Poland}




\date{\today}
\begin{abstract}
Moir\'e materials formed in two-dimensional semiconductor heterobilayers
are quantum simulators of Hubbard-like physics with unprecedented electron-density and interaction-strength tunability.
Compared to atomic scale Hubbard-like systems, electrons or holes in moir\'e materials
are less strongly attracted to their effective lattice sites because these
are defined by finite-depth potential extrema.
As a consequence, non-local interaction terms like interaction-assisted hopping and intersite-exchange are more  
relevant. We theoretically demonstrate the possibility of tuning the strength of 
these coupling constants to favor unusual states of matter, including spin liquids, insulating ferromagnets, and superconductors.



\end{abstract}

\pacs{Valid PACS appear here}
\maketitle

{\em Introduction:}---Moir\'e materials have emerged as an attractive controllable platform to simulate and explore quantum condensed 
matter \cite{Review1,Review2,FengchengHubbard, LiangFuTransfer,Mattia,MoireMIT}. The electronic structure of moir\'e materials is accurately described by continuum models with moir\'e 
spatial periodicity that can be engineered to yield Bloch bands with controllable width \cite{FengchengHubbard} and topology \cite{FengchengTopology,FengchengTMD1}. 
For moir\'e bilayers formed by transition metal dichalcogenides (TMD), electrons in the valence moir\'e band can experience triangular or honeycomb
lattice symmetry periodic potentials, depending on the TMD monolayer constituents, and 
the closest commensurate stacking arrangement.  For small twist angles 
the low-energy physics can correspondingly be described by either a single-band or a two-band model with a locked spin-valley 
pseudospin \cite{FengchengHubbard, LiangFuTransfer,Mattia,MoireMIT}. 
The emergent many-body physics, which is  extremely sensitive to 
the flat-band filling factor $\nu=N/N_{M}$,  can be modelled theoretically by adding electronic 
interactions to the continuum band model directly in momentum space \cite{FengchengTMD2,MoireMIT}
or by mapping the minibands to generalized Hubbard models.
(Here $N$ is the number of electrons or holes and $N_{M}$ is the number of moir\'e periods in the system).
Recent experiments in moir\'e TMD homobilayers and heterobilayers have exploited the possibility of tuning $\nu$ 
through large ranges with electrical gates,
discovering Mott \cite{HubbardCornell,Berkeley} and quantum anomalous Hall \cite{QAHCornell} 
insulating states at $\nu=1$ and generalized Wigner crystal states at several rational fractional fillings \cite{CornellWigner,CornellWigner2,CaliforniaWigner}.
The appearance of Wigner crystal states establishes the importance of long-range interactions in the 
many-body physics of semiconductor moir\'e materials, which are expected to enrich phase diagrams \cite{FengchengTMD2}.
\begin{figure}[h!]
\centering
\includegraphics[width=\linewidth]{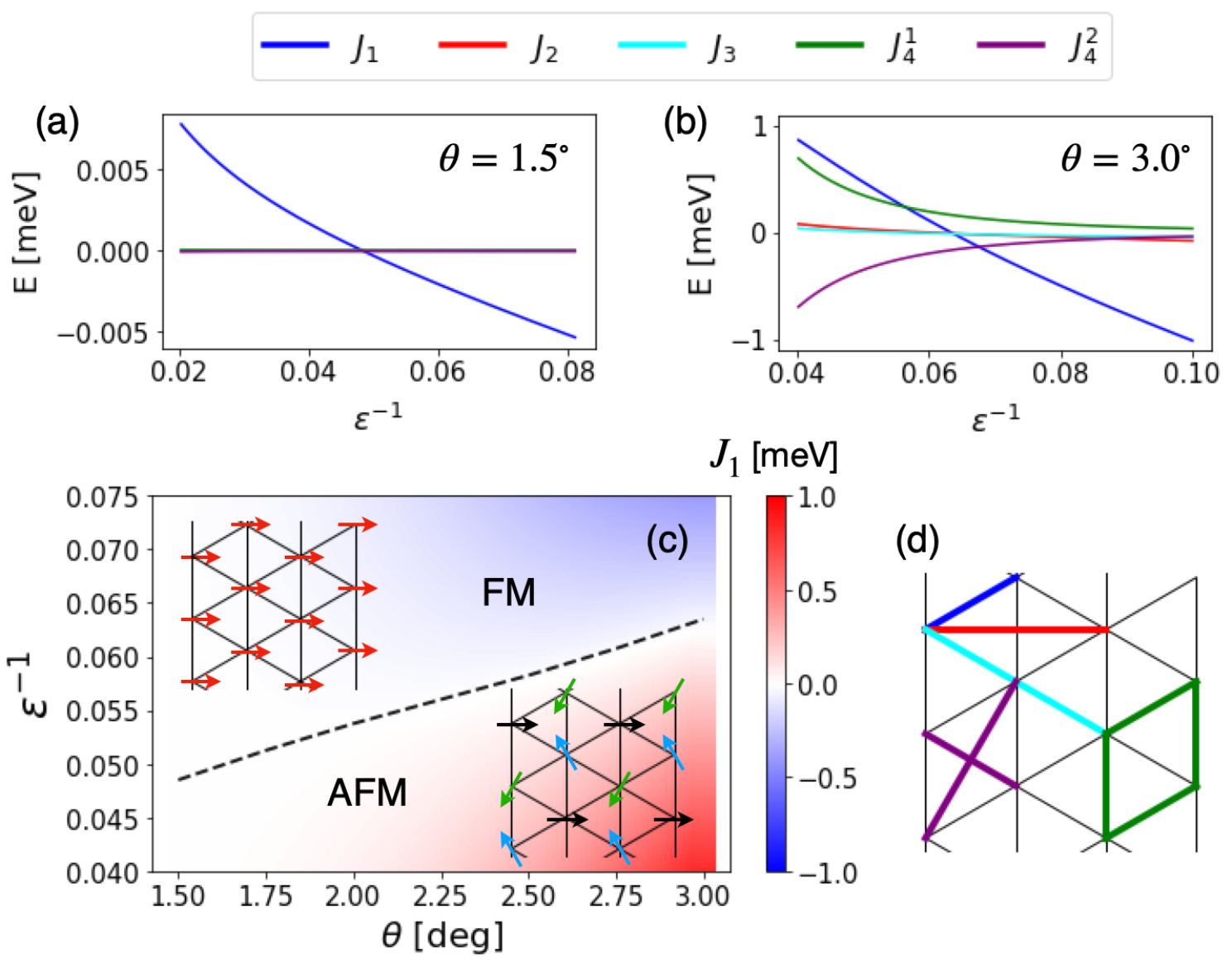}
\caption{Spin model coupling constants of the Mott insulator state at $\nu=1$ for a heterobilayer at small twist angle (long moir\'e length) (a) 
and at a larger twist angle (shorter moir\'e length) (b), as a function of interaction strength $\epsilon^{-1}$. (c) Phase diagram of a $\nu=1$ 
twisted heterobilayer {\it vs.} interaction strength $\epsilon^{-1}$ and twist angle $\theta$, indicating the antiferromagnet-ferromagnet transition line. Color scale shows the magnitude of the first-neighbor Heisenberg coupling, $J_1$. (d) Schematic illustrations of the neighbor configuration for the various coupling constants
presented in the upper panels (a) and (b). These calculations were performed for a modulation potential with $\psi=-94^{\circ}$ and $V_m=11$ meV (see main text).}
\label{fig:AntiferroFerrotransition}
\end{figure}

In this article we show that off-diagonal in site interactions,
often ignored in studies of Hubbard model physics, 
play a significant role in determining the ground state properties of semiconductor moir\'e materials.
Starting from continuum model Bloch states, we use a projection technique \cite{Cloizeaux1,Cloizeaux2,Vanderbilt} to obtain Wannier functions of holes localized on moir\'e superlattice sites. From these Wannier functions we calculate generalized Hubbard model parameters, that we use to derive a low-energy spin model description valid for strong interaction strengths at $\nu=1$. For small twist angles, or equivalently large moir\'e lattice constants, the Wannier orbitals are well approximated by the eigenstates of a harmonic potential and therefore an on-site Hubbard model description is justified. Decreases
in the moir\'e lattice constant or the modulation potential strength lead to overlaps between the tails of Wannier functions localized 
on nearest-neighbor lattice sites (See supplemental material \cite{Supplemental}).  When significant, the overlap gives rise to enhanced non-local interaction terms.
In exploring their influence, we have focused on the spin-physics of Mott insulator states at $\nu=1$.
Our main results are presented in Fig. \ref{fig:AntiferroFerrotransition}. In Fig. \ref{fig:AntiferroFerrotransition}(a)-(b) we show Heisenberg model spin coupling constants
for a small twist angle with well-localized Wannier orbitals, and for a larger twist angle with  
significant Wannier function overlap between neighbors. 
As illustrated in \ref{fig:AntiferroFerrotransition}(c), we find that the nearest-neighbor interaction $J_1$ 
changes sign as a function of twist angle and background dielectric screening, 
indicating the possibility of controlled tuning between antiferromagnetic and ferromagnetic states.  
We confirm this transition by finite size exact diagonalization calculations. 
Hartree-Fock analyses of heterobilayers \cite{NaichaoAllan,MakQAH} and homobilayers \cite{FengchengTMD1,ColumbiaCano} have also identified a ferromagnetic phase as a candidate ground state at $\nu=1$ when dielectric screening is weak. For larger twist angles, which are more relevant experimentally, contributions of other two- and four-spin terms become important in spin model descriptions of TMD moir\'e materials. Our findings suggest strategies to create unusual states, including  
ferromagnetic insulators, spin liquids, and superconductors.
\vspace{0.5cm}

{\em Generalized Hubbard model for moir\'e TMDs:}--- We limit our attention to TMD heterobilayers 
that form triangular moir\'e superlattices and therefore permit a single-band low-energy description with trivial topology. 
Assuming a smooth potential limit \cite{FengchengHubbard}, the continuum model that describes the bilayer's electronic structure depends only on
the moir\'e lattice constant $a_M$, the modulation potential strength $V_m$, and a single potential-shape parameter $\psi$
(For details on the continuum model see the supplemental material \cite{Supplemental}). The continuum model can be mapped to a real space lattice model, whose Hamiltonian is written in the most general way as
\begin{align}
\label{ElectronInteractions}
    H=-\sum_{i,j,\sigma}  &t_{ij}\, c^{\dagger}_{i,\sigma} c_{j, \sigma}\nonumber \\
    &+\frac{1}{2}\sum_{\substack{i,j,k,l\\
    \sigma \sigma^{\prime}}} V_{ijkl}^{\sigma, \sigma^{\prime}} c^{\dagger}_{i, \sigma}c^{\dagger}_{j,\sigma^{\prime}}c_{l, \sigma^{\prime}} c_{k, \sigma},
\end{align}
where $c^{\dagger}_{i,\sigma} (c_{i,\sigma})$ creates (destroys) an electron at site $i$ in valley $\sigma$; $i,j,k,l$ are site labels, $t_{ij}$ stands for the hopping integral between sites $i$ and $j$, and $V_{ijkl}^{\sigma \sigma^{\prime}}$ is a two-particle matrix element
\begin{align}
    \label{CoulombReal}
    V_{i,j,k,l}^{\sigma,\sigma^{\prime}}=\braket{{\bf R}_i,{\bf R}_j|V|{\bf R}_k,{\bf R}_l},
\end{align}
with ${\bf R}_i$ the moir\'e lattice site positions. The Coulomb long-range interaction is given by $V=e^2/\epsilon|{\bf r}_1-{\bf r}_2|$  and $\epsilon^{-1}$ is the system's dielectric screening from the surrounding environment, which determines the interaction strength. Since $V_{i,j,k,l}^{\sigma,\sigma^{\prime}}$ is invariant under 
global translations, we can choose ${\bf R}_i=0$. The largest matrix elements are the on-site interactions $U_0=\braket{{\bf 0},{\bf 0}|V|{\bf 0},{\bf 0}}$, and two-center integrals involving
sites ${\bf 0}$ and ${\bf R}$. The latter include the nearest-neighbor direct interaction $U_1=\braket{{\bf 0},{\bf R}|V|{\bf 0},{\bf R}}$, intersite-exchange $X_1=\braket{{\bf 0},{\bf R}|V|{\bf R},{\bf 0}}$, assisted hopping $A_1=\braket{{\bf 0},{\bf 0}|V|{\bf 0},{\bf R}}$ and pair-hopping $P_1=\braket{{\bf 0},{\bf 0}|V|{\bf R},{\bf R}}$ matrix elements. 

For single-particle potentials that are strongly attractive on lattice sites, like those of atomic-scale ionic crystals, 
Wannier-functions are well localized, and non-local interactions that require overlap between distinct Wannier functions are usually negligible. In the intermediate case of $d$-band electrons in an elemental transition metal crystal
Hubbard estimated that $U_0 \sim 20$ eV, $U_1\sim 6$ eV, $A\sim 0.5$ eV and $X,P\sim 1/40$ eV \cite{HubbardUterms}. 
Because nearest-neighbor interaction terms can be reduced by screening, it is sometimes justified to 
retain only $U_0$, yielding the standard on-site Hubbard model. 
In general, a less attractive potential has more extended Wannier functions, 
modifying the relationship between the various interaction terms.
Non-local interactions have been considered previously in extended Hubbard model theories 
of polyacetyline \cite{KivelsonSSH,TinkaGammel1,TinkaGammel2}, 
where they enhance dimerization, and can produce a ferromagnetic phase but only in parameter ranges that 
appear to be unphysical.  Because the assisted hopping interaction may acquire a large 
multiplicative factor related to lattice geometry, it can play a significant role even when much smaller than $U_0$,
potentially causing pairing and leading to superconductivity \cite{HirschMarsiglio, Hirsch}.
In the following we address the importance of non-local terms in twisted TMD heterobilayers,
concentrating on their role in determining $\nu=1$ ground state properties.

From eigenvectors and eigenvalues of the continuum model's topmost band we obtain Wannier functions localized at moir\'e lattice sites and evaluate extended Hubbard model parameters $t_{i,j}$ and $V_{i,j,k,l}$, shown  as  lines  with  dots  in  Fig. \ref{fig:Uterms}, as described in \cite{Supplemental}. As a consistency check, we compare our extended Hubbard model parameters with the ones obtained in the regime of large $a_M$, where the modulation potential minima can be approximated by a set of harmonic potentials centered on moir\'e lattice sites \cite{FengchengHubbard} and analytic control is possible. In this limit, the Wannier functions are 
\begin{align}
    \psi_{{\bf R}}({\bf r})=\left(\frac{1}{\pi a_W^2}\right)^{1/2}\text{exp }\left[- \frac{({\bf r-R})^2}{2a_W^2}\right],
\end{align}
where $a_W=\kappa^{1/4}\sqrt{a_M}$ is the Wannier function width and $\kappa=\hbar^2/(16 \pi^2 V_m m^* \cos(120^{\circ}+\psi) )$ varies inversely 
with modulation potential strength. In this approximation $a_M\sim a_0/\theta$, with $a_0$ the active layer's lattice constant. We find that the near-neighbor hopping amplitude is
\begin{align}
\label{hop}
    t&=\frac{\hbar^2}{2m^*a_W^2}\left(\frac{a_M^2}{4 a_W^2}-1 \right)\text{exp }\left[-\frac{a_M^2}{4 a_W^2} \right]\nonumber \\
    &=\frac{\hbar^2}{2m^*\kappa^{1/2}}\left(\frac{1}{4 \kappa^{1/2}}-\frac{\theta}{a_0} \right)\,\text{exp}\left[-\frac{a_0}{4\kappa^{1/2}\theta} \right],
\end{align}
while the most significant interaction matrix elements are
\begin{align}
    U_0=\frac{\pi^{1/2}e^2}{\sqrt{2}\,\epsilon \, a_W}\sim \sqrt{\theta}&,
\end{align} 
\begin{align}
    U_1=\frac{2 e^2 I_1}{\sqrt{\pi}\,\epsilon \, a_W}\sim \sqrt{\theta}\,&I_1,
\end{align}
\begin{align}
    A_1=\frac{I_2}{I_1}U_1 ~ \text{exp}\left[-\frac{a_M^2}{4a_W^2} \right]\sim \sqrt{\theta}\,I_2\, &\text{exp}\left[-\frac{a_0}{4\kappa^{1/2}\theta} \right],
\end{align}
\begin{align}
    X_1=P_1=U_0~\text{exp}\left[ -\frac{a_M^2}{2 a_W^2}\right]\sim \sqrt{\theta}~&\text{exp}\left[-\frac{a_0}{2\kappa^{1/2}\theta} \right],
    \label{Assis}
\end{align}
with $I_1$ and $I_2$ integrals given in the supplemental material \cite{Supplemental}.

A comparison between the analytical expressions given by Eqs. (\ref{hop})-(\ref{Assis}) and the results for extended Hubbard model parameters obtained from numerical calculations as a function of twist angle is provided in Fig. \ref{fig:Uterms}. For small twist angles we see good agreement, as expected, while for larger twist angles the harmonic approximation underestimates Wannier function tails, and therefore non-local interaction
strengths.  Interestingly, we see in Fig. \ref{fig:Uterms}(c) that the non-local exchange interaction $X_1$ 
increases significantly with twist angle, and in Fig. \ref{fig:Uterms}(d) that there is a range of angles for which 
the assisted hopping amplitude $A_1$ becomes negative. These qualitative differences between the harmonic potential approximation and exact results are expected 
since the lattice potentials in the former model have unbounded strength, whereas the actual potential is bounded, causing that for $\theta \gtrsim 2.0^{\circ}$ the Wannier functions are more extended and acquire negative tails \cite{Supplemental}. 

\begin{figure}[h!]
\centering
\includegraphics[width=0.85\linewidth]{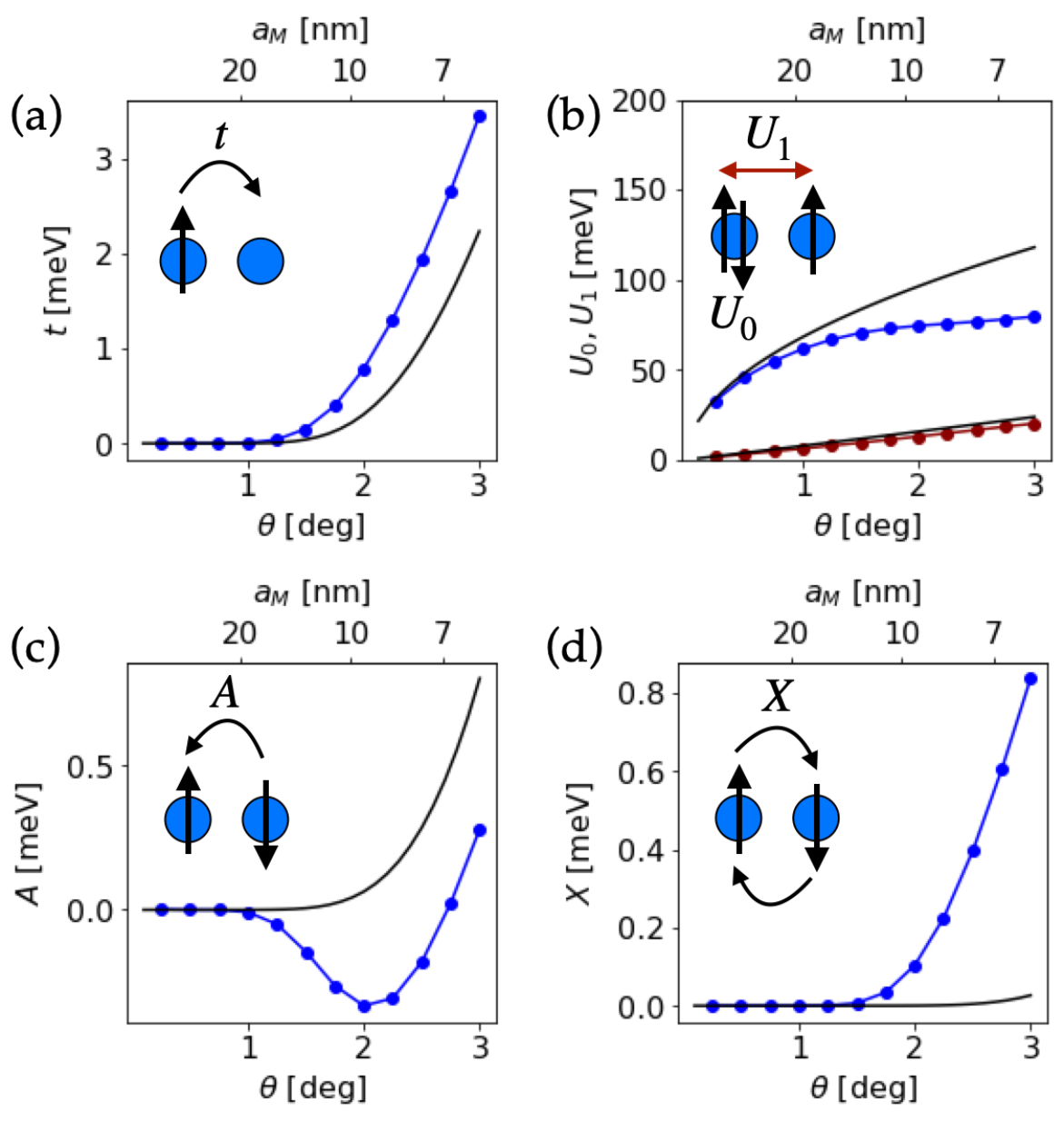}
\caption{Comparison between harmonic approximation (solid black lines) and exact continuum model results (lines with dots) for (a) hopping, (b) on-site (blue) and first-neighbor (dark red) interactions, (c)  assisted hopping, and (d) intersite exchange interactions. Interaction parameters are plotted {\it vs} twist angle (bottom axes) and {\it vs} moir\'e length (top axes). The insets provide schematic illustrations of each process. These calculations are for $V_m=11$ meV, $\psi=-94^{\circ}$, and $\epsilon=10$.}
\label{fig:Uterms}
\end{figure}

{\em Effective spin model}---
To illustrate the qualitative impact of non-local interactions on moir\'e Hubbard physics, we focus on the spin-physics of 
the Mott insulator states at $\nu=1$.  The charge gap of the Mott insulators is set by the  
$U_0$ energy scale that makes double-occupancy of any lattice site energetically unfavorable. 
When $U_0$ is larger than all other energy scales, the Hubbard spectrum separates into two branches, 
an upper branch with a large double occupation weight, and a low-energy branch in which charge is approximately frozen and is described by a spin Hamiltonian 
\begin{align}
     H_{\text{eff}}&=J_1\sum_{\langle i,j\rangle}{\bf S}_i\cdot{\bf S}_j + J_2\sum_{\langle\langle i,j \rangle\rangle}{\bf S}_i\cdot{\bf S}_j+J_3\sum_{\braket{\braket{\braket{i,j}}}}{\bf S}_i\cdot{\bf S}_j\nonumber \\
    &\quad +\sum_{\ring}J_4^1\left[({\bf S}_i\cdot{\bf S}_j)({\bf S}_k\cdot{\bf S}_l)+({\bf S}_i\cdot{\bf S}_l)({\bf S}_j\cdot{\bf S}_k)\right]\nonumber\\
    &\qquad \qquad +J_4^2({\bf S}_i\cdot{\bf S}_k)({\bf S}_j\cdot{\bf S}_l),    
    \label{EffectiveSpin}
\end{align}
where the ${\bf S}_{i}$ are spin operators and the summations are over first nearest-neighbors, 
second nearest-neighbors, third nearest-neighbors and ring clusters, respectively. 
The coupling constants of the spin model can be expressed in terms of the real space Coulomb matrix elements 
by applying a cluster perturbation expansion \cite{AllanLargeU} or equivalently a Schrieffer-Wolff transformation \cite{AllanSW} to the Hamiltonian in 
Eq. \eqref{ElectronInteractions}, as detailed in the supplemental material \cite{Supplemental}. We show that the dominant near-neighbor coupling constant $J_1 \approx  4 (t_1-A_1)^2/(U_0-\nobreak U_1) - 2 X_1$ has independent contributions from two 
different mechanisms, an antiferromagnetic super-exchange contribution that is inversely proportional to interaction
strength and a ferromagnetic direct exchange contribution that is proportional to interaction strength.
Because the two contributions respond oppositely to changes in interaction strength,
the one that dominates can be changed by controlling the dielectric constant $\epsilon$ of the surrounding material.
Typical results for the dependence of spin-model coupling constants on $\epsilon^{-1}$ are shown in Figs. \ref{fig:AntiferroFerrotransition}(a) 
and \ref{fig:AntiferroFerrotransition}(b) for angles $\theta=1.5^{\circ}$ and  $\theta=3.0^{\circ}$ respectively. 

\begin{figure}[h!]
\centering
\includegraphics[width=\linewidth]{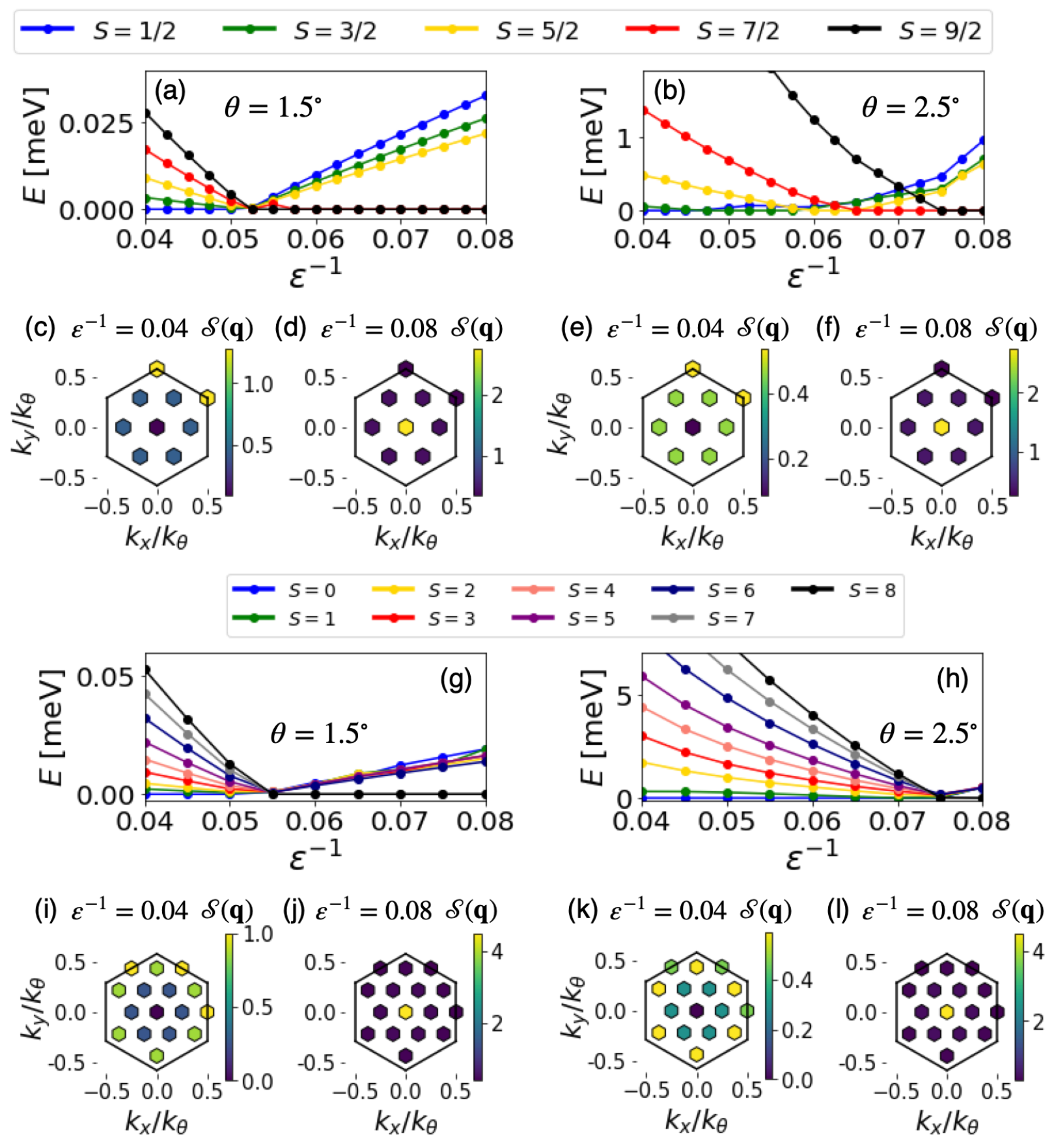}
\caption{Transition from antiferromagnet to ferromagnet as seen by exact diagonalization of the continuum model. 
The evolution of the lower Hubbard band {\it vs.} $\epsilon^{-1}$ for (a) $\theta=1.5^{\circ}$ and (b) $\theta=2.5^{\circ}$. 
Evolution of the lowest energy state for each $S$ {\it vs.} $\epsilon^{-1}$ for (c) $\theta=1.5^{\circ}$ and (d) $\theta=2.5^{\circ}$. Spin structure factors for twist angle $\theta=1.5^{\circ}$ in the antiferromagnetic (e) and ferromagnetic (f) phases and for twist angle $\theta=2.5^{\circ}$ in the antiferromagnetic (g) and ferromagnetic (h) phases. Axes in structure factor plots are normalized by $k_{\theta}=4\pi/\sqrt{3}a_M$, the length of the moir\'e reciprocal lattice vectors.}
\label{fig:ED_Structure}
\end{figure}

In Fig. \ref{fig:AntiferroFerrotransition}(a) we see that for small angles or long moir\'e periods, $J_1$ is the dominant coupling constant.
The many-body ground state of the system is expected to be antiferromagnetic for $J_1>0$ and ferromagnetic for $J_1<0$. To demonstrate this behavior explicitly, we calculate the full low-energy spectrum of the TMD bilayer by 
finite-size exact diagonalization of the continuum model. Performing ED directly in momentum space allows us to include all long-range interactions. The evolution of the lowest eigenvalue with total spin quantum number $S$ for $\theta=1.5^{\circ}$ and $\theta=2.5^{\circ}$, with respect to $\epsilon^{-1}$, is plotted in Fig. \ref{fig:ED_Structure}(a) and (b) for $N=9$ and in Fig. \ref{fig:ED_Structure}(g) and (h) for $N=16$. From these results we see that for the smaller angle the ground state is a singlet when $J_1>0$, as expected for an antiferromagnetic state and that for the region where $J_1<0$ the ground state is a ferromagnet. 
The spin structure factors calculated in the antiferromagnetic phase, shown in Fig.  \ref{fig:ED_Structure}(c),(i), show peaks at the corners of the Brillouin zone, indicating a 3-sublattice state, while the structure factors in the ferromagnetic phase, shown in Fig.  \ref{fig:ED_Structure}(d),(j), have a peak at $\gamma$, characteristic of a ferromagnetic state.
At larger twist angles the harmonic approximation is not accurate and nearest-neighbor coupling $J_1$ is less dominant. In this case we also have a ferromagnetic insulating ground state for large $\epsilon^{-1}$ and an antiferromagnetic ground state for small $\epsilon^{-1}$ for both system sizes, as can be seen from total spin $S$ plots, Fig. \ref{fig:ED_Structure} (b),(h) and structure factors, Fig. \ref{fig:ED_Structure}(e),(f),(k),(l). 
The region near where $J_1$ changes sign is now more complex, as can be observed from our finite-size calculations. Although our 
ED calculations cannot determine the thermodynamic limit ground state in this regime, it is clear that exotic spin-states are 
likely to be abundant close to the antiferromagnet-ferromagnet transition. Ring-exchange terms $J_4^1$ and $J_4^2$ become significant and may favor spin liquid ground states \cite{Zaletel,Cookmeyer} and the contributions from $J_2$ and $J_3$ also suggest exotic spin states. In Fig.~\ref{fig:AntiferroFerrotransition}(b), as interaction strength $\epsilon^{-1}$ increases, the superexchange couplings $J_3,\ J_2,\ J_1$ change sign from positive to negative sequentially. In the region where $J_3<0$ but $J_1>J_2\gtrsim0$, there is bound to be a point where $-J_1/J_3 = 9$. Close to that point, another antiferromagnetic spin configuration, the stripe state \cite{JolicoeurPhysRevB1990,NaichaoAllan}, has a very similar classical energy to the 3-sublattice state, making quantum fluctuations important in determining the ground state.

{\em Discussion:}---    We have shown that non-local interaction terms can be sizable in semiconductor moir\'e materials and that they can have an important influence on electronic properties, giving rise to moir\'e Mott-Hubbard ferromagnets, not expected in other systems described by Hubbard models with only local interaction terms. Non-local interactions become more prominent at larger twist angles and at weaker moir\'e modulation, where a harmonic expansion of the modulation potential near its minima fails to describe the band Wannier functions (See Fig. \ref{fig:Uterms}), justifying the methodology employed here.
In the case of the Mott insulator states that appear at moir\'e filling factor $\nu=1$, non-local exchange supplies a ferromagnetic contribution to the near-neighbor interaction between spins that is comparable in strength to the antiferromagnetic superexchange contribution, making sign changes in the total interaction common over typical ranges of experimental parameters. In particular, current WSe$_2$/WS$_2$ samples with $a_M\sim 8$ nm \cite{Berkeley,CornellWigner,CornellWigner2,CaliforniaWigner} appear on the antiferromagnetic side of the phase boundary and the competition with ferromagnetism can be tuned {\em in situ} by varying the moir\'e modulation strength, which mainly influences $t_1$ - using gate electric fields \cite{ExperimentMIT1,ExperimentMIT2} or pressure \cite{Dean1, Dean2} - or background screening of electronic interactions, providing a promising framework to confirm the phase transition in the future. Our findings establish a strategy for engineering strongly frustrated spin-Hamiltonians that are likely to host exotic spin states.

In our explicit calculations we have considered only the case of wavevector and frequency independent background screening of the type produced by a thick surrounding dielectric, but more general situations are also relevant.
(We have focused on a range of $\epsilon^{-1}$ values that is smaller than what would be produced by screening by a surrounding 
hBN dielectric alone ($\epsilon^{-1} \sim 0.2$), in anticipation of additional screening 
by conducting gates and by virtual transitions between flat and energetically remote moir\'e minibands). Similar conclusions apply to more complex moir\'e material states. For example,
it has been established experimentally that non-near-neighbor local interaction terms $U_n$ are important in moir\'e TMD systems,
and that they give rise to insulating Wigner crystal states at many fractional values of $\nu$ \cite{CornellWigner,CornellWigner2,CaliforniaWigner}, (Presumably these Wigner crystal states would also appear in real crystals if it were possible to change the electron density without
introducing disorder). The generalized Wigner crystal states also have low-energy spin-sectors whose interactions are more complex than those of the $\nu=1$ case considered here but will have coupling constants that are tunable in sign due to the  
competition between direct and superexchange spin interactions, determining their magnetic properties.  Separately, in honeycomb lattice moir\'e materials \cite{FengchengTopology,FengchengTMD1,QAHCornell,MakQAH,BartHofstadter} spin-physics can be entangled with topologically non-trivial band-mixing, adding another wrinkle to the low-energy physics, opening the possibility of realizing fractional Chern insulators. 
Finally, we remark that we have focused here on the near-neighbor exchange non-local interaction because it is
particularly important at $\nu=1$.  Other non-local interactions may play a more prominent role at metallic filling factors.   
For example, it has been proposed \cite{SuperconductivityFu,HirschMarsiglio} that assisted hopping 
can trigger superconductivity. All these issues 
deserve attention in future work. 

The authors acknowledge helpful interactions with Kin Fai Mak and Jie Shan. We also thank Johannes Motruk for a careful examination of the spin model expressions.  We acknowledge the Texas Advanced Computing Center (TACC) at The University of Texas at Austin for providing high-performance computer resources. PP acknowledges support from the Polish National Science Centre based on Decision No. 2021/41/B/ST3/03322. This work was supported by the U.S. Department of Energy, Office of Science, Basic Energy Sciences, under Award $\#$ DE-SC0022106. 

\bibliography{refs}

\onecolumngrid
\section*{Continuum model and Wannier functions for TMD bilayers}
Due to layer asymmetry in twisted TMD heterobilayers, electrons or holes populate the valence band of only one of the layers (the active layer). The presence of the other layer generates a potential with the moir\'e periodicity that affects electrons or holes in the valence band. Valley degeneracy in these systems is lifted by spin-valley locking, meaning that we can consider only one valley (or spin) which is related to the other by time-reversal symmetry. The valley-projected continuum Hamiltonian for TMD moir\'e heterobilayers is given by
\begin{align}
    \mathcal{H}=-\frac{\hbar^2}{2m^*}{\bf k}^2+\Delta({\bf r}),
    \label{MoireHamiltonian}
\end{align}
where $m^*$ is the effective mass of charge carriers in the valence band of the active layer. For calculations presented in this work we have taken $m^*=0.35\, m_0$, assuming that the active layer is $\text{WSe}_2$. The modulation potential is assumed to be a smooth function with the superlattice periodicity that can be approximated in a Fourier expansion as \cite{FengchengHubbard}
\begin{align}
\Delta({\bf r})=2V_m\sum_{j=1,3,5}\cos({\bf b}_j\cdot{\bf r}+\psi)   
\end{align}
with ${\bf{b}}_j=4\pi/\sqrt{3}a_M\left(\cos\left(\pi j/3\right),\sin\left(\pi j/3\right)\right)$, belonging to the first shell of reciprocal lattice vectors and ($V_m$,\,$\psi$) two parameters that determine the strength of the potential and the location of its minima, respectively. The values of the potential parameters are obtained from {\it ab initio} calculations and vary between different authors, nevertheless applying an out-of-plane external field or pressure can effectively vary the strength of the modulation potential. For that reason we take $V_m$ as a controllable parameter in our study. It has been established that this model yields a topmost isolated flat band that can be mapped to a triangular lattice for $\psi=-94^{\circ}$ \cite{FengchengHubbard}, the value we fix for our calculations. A basis of Bloch functions that diagonalizes the moir\'e Hamiltonian \eqref{MoireHamiltonian} is
\begin{align}
    \ket{{\bf k},n}=\sum_{\bf G}z_{{\bf k+G}}^n\ket{{\bf k+G}},
\end{align}
where $\ket{{\bf k+G}}$ are plane waves and $n$ is the band index. Starting from those states, Wannier functions for the flat band (thus band index is omitted) are obtained
\begin{align}
     \ket{{\bf R}}=\frac{1}{\sqrt{N_M}}\sum_{\bf k}e^{-i\,{\bf k}\cdot{\bf R}}\ket{{\bf k}},
\end{align}
where ${\bf R}$ are triangular moir\'e lattice sites and positions of Wannier centers and $N_M$ is the number of unit cells in the system, or equivalently, the number of moir\'e sites. Coulomb elements $V_{i,j,k,l}^{\sigma,\sigma^{\prime}}$, as defined in the main text, can be obtained via Eq. \eqref{CoulombReal} by directly calculating the matrix elements involving four sites in real space once the Wannier functions are obtained \cite{FengchengTMD2}. An equivalent way is to obtain all Coulomb elements in momentum space $\braket{{\bf k}_i,\sigma;{\bf k}_j,\sigma^{\prime}|V|{\bf k}_k,\sigma;{\bf k}_l,\sigma^{\prime}}$ and take a Fourier transform
\begin{align}
\label{Coulomb}
    V^{\sigma,\sigma^{\prime}}_{i,j,k,l}=\frac{1}{N_M^2}\sum_{\substack{{\bf k}_i,{\bf k}_j\\{\bf k}_k,{\bf k}_l}}e^{i ({\bf k}_i\cdot{\bf R}_i+{\bf k}_j\cdot{\bf R}_j-{\bf k}_k\cdot{\bf R}_k-{\bf k}_l\cdot{\bf R}_l)} \braket{{\bf k}_i,{\bf k}_j|V|{\bf k}_k,{\bf k}_l},
\end{align}
where the spin labels are omitted for shorthand. There is a phase freedom in the Bloch states which determines the localization of the Wannier functions. Since our goal is to accurately describe the bilayer in terms of a real space Hamiltonian, we choose a gauge which yields localized and real Wannier functions \cite{Vanderbilt}. This choice ensures that Coulomb elements between Wannier functions whose centers are separated by distances larger than ${\bf R}$ decrease rapidly, yielding a finite set of relevant parameters that determine the many-body physics. For small twist angles, Wannier functions are localized in real space and extended in momentum space, meaning we require a larger basis of reciprocal lattice vectors ${\bf G}$ in order to have well-converged results. We verified that taking 127 ${\bf G}$-vectors, corresponding to six shells in reciprocal space, is sufficient in the range of twist angles considered in this paper.

\begin{figure}[h!]
\centering
\includegraphics[width=0.9\linewidth]{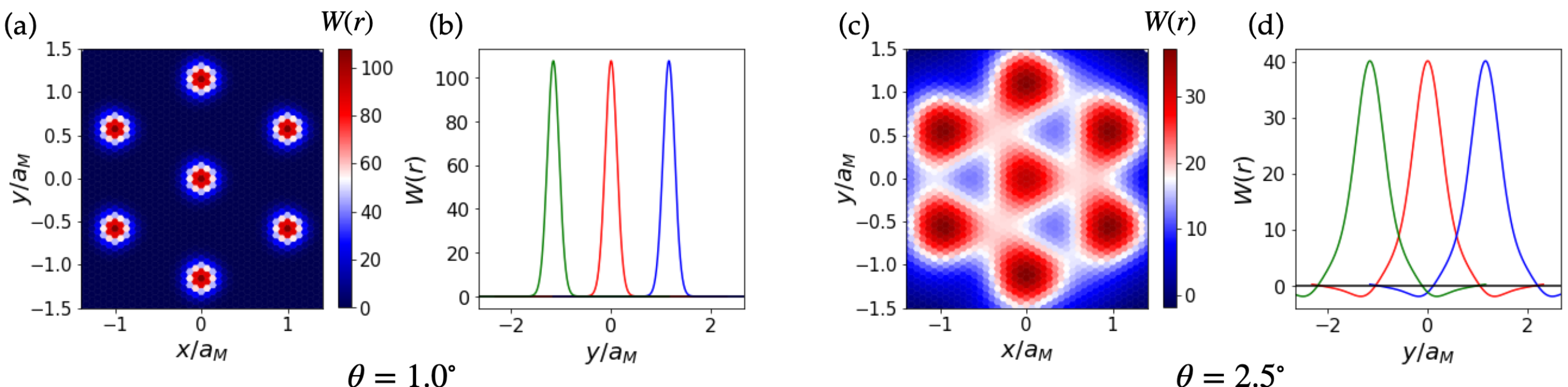}
\caption{Wannier functions located at triangular moir\'e sites, obtained from the continuum model for $\theta = 1.0^{\circ}$ (a) and $\theta = 2.5^{\circ}$ (c), with $V_m=11$ meV and $\psi=-94^{\circ}$. Line-cuts along $x=0$ of (a) and (c) are shown in (b) and (d) respectively.}
\label{fig:Wanniers}
\end{figure}

In Fig. \ref{fig:Wanniers} we show Wannier functions for two twist angles, $\theta = 1.0^{\circ}$ and $\theta = 2.5^{\circ}$. For the larger angle one can clearly see tails of Wannier functions with negative values around positions of neighboring lattice sites. The exchange interaction can be understood as a measure of how much neighboring Wannier functions overlap, hence the increased values for larger angles due to tails. In the case of assisted hopping, the product in the two-site integral $\braket{{\bf 0},{\bf 0}|V|{\bf 0},{\bf R}}$ can become negative if the tail corresponding to $\ket{\bf R}$ overlaps with the peak of the other three Wannier functions.
\section*{Exact diagonalization calculations in momentum space}
Interactions can be added to the continuum single-particle model by projecting them to the flat band, when it is isolated from remote bands, yielding the following Hamiltonian
\begin{align}
\label{ManyBodyHamiltonian}
    H&=\sum_{{\bf k}, \sigma} \epsilon_{{\bf k}, \sigma}~ c^{\dagger}_{{\bf k},\sigma} c_{{\bf k}, \sigma}+\frac{1}{2}\sum_{\substack{{\bf k}_1^{\prime},{\bf k}_2^{\prime}\\{\bf k}_1,{\bf k}_2}}\sum_{\sigma \sigma^{\prime}} V_{k_1^{\prime} k_2^{\prime} k_1 k_2}^{\sigma \sigma^{\prime}} c^{\dagger}_{{\bf k}_1^{\prime}, \sigma}c^{\dagger}_{{\bf k}_2^{\prime},\sigma^{\prime}}c_{{\bf k}_2, \sigma^{\prime}} c_{{\bf k}_1, \sigma},
\end{align}
where $c^{\dagger}_{{\bf k},\sigma} (c_{{\bf k},\sigma})$ creates (destroys) a hole with momentum ${\bf k}$ in
valley $\sigma$; ${\bf k}_1,{\bf k}_2,{\bf k}'_1,{\bf k}'_2$ are momentum labels, $\epsilon_{{\bf k},\sigma}$ is a flat valence band single-particle energy obtained from the continuum model, 
and $V_{k_1^{\prime} k_2^{\prime} k_1 k_2}^{\sigma \sigma^{\prime}}$ is a two-particle matrix element 
\begin{align}
    V_{k_1^{\prime} k_2^{\prime} k_1 k_2}^{\sigma \sigma^{\prime}}=\langle {\bf k}_1^{\prime},\sigma;{\bf k}_2^{\prime},\sigma^{\prime}|V|{\bf k}_1,\sigma;{\bf k}_2,\sigma^{\prime} \rangle.
    \label{CoulombElementMomentum}
\end{align}
These are the matrix elements used to calculate the real space Coulomb elements via Eq. \eqref{Coulomb}. Note also that Eq. \eqref{ManyBodyHamiltonian} corresponds to the momentum space representation of the Hamiltonian \eqref{ElectronInteractions} presented in the main text. When the latter Hamiltonian is approximated to a Hubbard-like model some of the interaction terms will be neglected. By considering interactions directly in momentum space we are including all short and long-range interactions, as well as non-local interaction terms, therefore we are not neglecting any contributions that could be relevant for the low-energy many-body physics.

\begin{figure}[h!]
\centering
\includegraphics[width=0.9\linewidth]{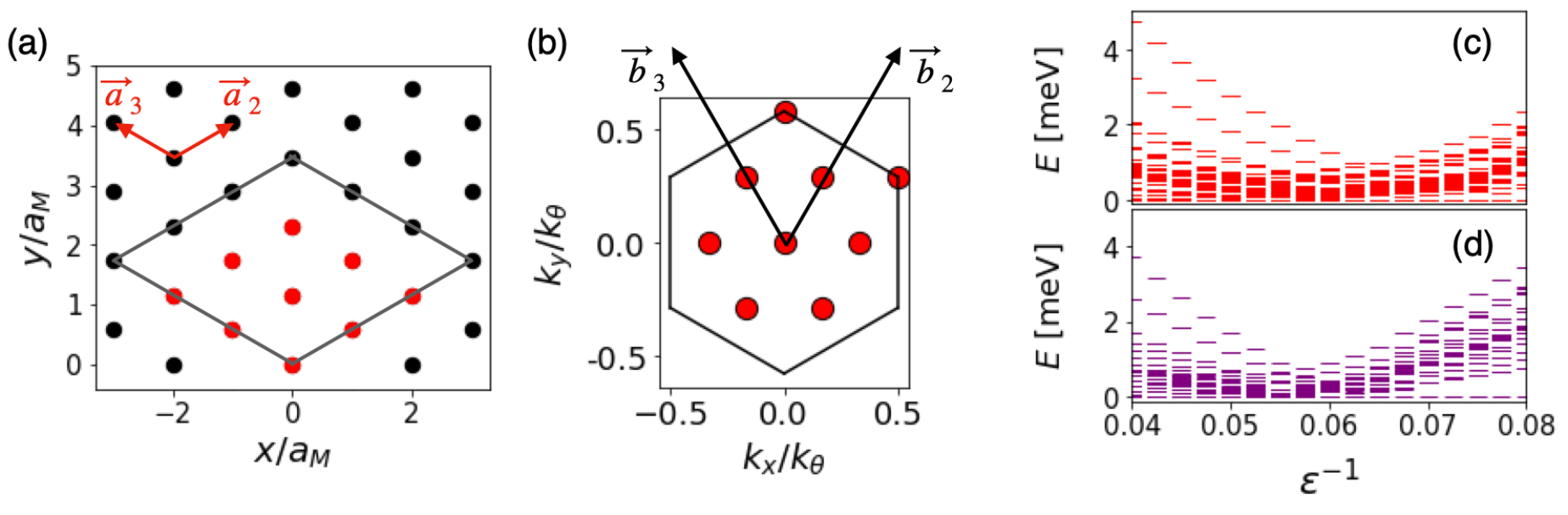}
\caption{(a) Real space supercell corresponding to the finite momentum mesh (b) used to diagonalize the many-body Hamiltonian, Eq. \eqref{ManyBodyHamiltonian}, with $N_2=N_3=3$. The mesh with $N_2=N_3=4$ is constructed in an analogous way. (c) Spectrum obtained by diagonalizing Eq. \eqref{ManyBodyHamiltonian} in the momentum space mesh shown in (b) for $\theta=2.5^{\circ}$, $V_m=11$ meV and $\psi=-94^{\circ}$, as a function of $\epsilon^{-1}$. (d) Exact diagonalization spectrum of an extended Hubbard model in real space with non-local and long-range interactions in the supercell shown in (a), as a function of $\epsilon^{-1}$.} 
\label{fig:EDmesh}
\end{figure}

We diagonalize the Hamiltonian \eqref{ManyBodyHamiltonian} in momentum space meshes of size $N=N_2\times N_3$  that apply periodic boundary conditions across supercells in real space. Examples for the supercell and the momentum mesh corresponding to $N_2=N_3=3$ are shown in  Fig. \ref{fig:EDmesh}(a) and (b), respectively. The points forming the momentum mesh are of the form ${\bf k}=n_2\, {\bf b_2}/N_2+n_3\, {\bf b_3}/N_3$, where ${\bf b_i}$ are the reciprocal lattice vectors, the corresponding plaquette in real space is spanned by the vectors $N_2\,{\bf a_2}$ and $N_3\,{\bf a_3}$, where ${\bf a_i}$ are the real space lattice vectors. The evolution of the lowest $2^N$ states (the spin sector) of the many-body spectrum resulting from momentum space ED is illustrated for $\theta=2.5^{\circ}$ in Fig. \ref{fig:EDmesh}(c), as a function of interaction strength. Additionally, we calculate the many-body spectrum of a real space Hubbard model including interaction parameters $t_1, t_2, t_3, U_0, U_1, U_2, U_3, X_1$ and $A_1$, obtained from the continuum model as described in the previous section, in a plaquette with N=9 moir\'e sites at half-filling. The resulting spectrum, as a function of interaction strength, is shown in Fig. \ref{fig:EDmesh}(d), showing good agreement with momentum space ED, Fig. \ref{fig:EDmesh}(c). This indicates that the description of the heterobilayer system by an extended Hubbard model is faithful only if long-range and non-local interactions are included, as expected.

Because the many-body Hamiltonian is invariant under translation and periodic boundary conditions have been applied, single particle operator expectation values like charge density and spin density will always be independent of position. In order to capture broken translation symmetry in charge or spin density wave states, if they occur, it is necessary to evaluate two-body correlation functions like the spin structure factors 
\begin{align}
    \mathcal{S}({\bf q})=\braket{S({\bf q})S({\bf -q})}=\sum_{{\bf R}}e^{-i{\bf q}\cdot{\bf R}}\langle \bf{S}(\bf{0})\cdot \bf{S}(\bf{R})\rangle,
\end{align}
shown in Fig. \ref{fig:ED_Structure} in the main text. Broken translational symmetry is signalled by a large value of $\mathcal{S}({\bf q})$ at a non-zero value of ${\bf q}$.
\section*{Extrapolation of Coulomb matrix elements}
\begin{figure}[h!]
\centering
\includegraphics[width=0.6\linewidth]{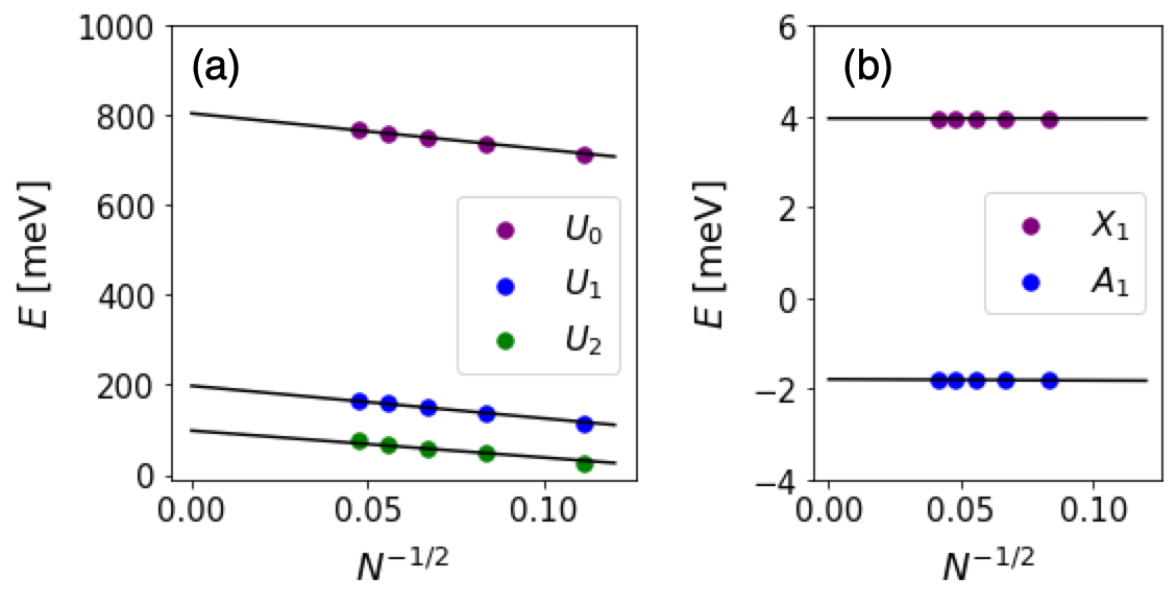}
\caption{Values of real space Coulomb elements calculated for Brillouin zone grids of different sizes as a function of $N^{1/2}$. (a) On-site, first-neighbor and second-neighbor interactions, (b) exchange and assisted hopping.}
\label{fig:Long}
\end{figure}

We estimate how finite size effects affect our results by calculating Coulomb elements for Brillouin zone grids of sizes $N=144,225,324,441$ and $576$ and extrapolating them to the thermodynamic limit, as shown in Fig. \ref{fig:Long}(a),(b). We see that $U_0$ for $N=441$ differs from its thermodynamic value by less than $5\%$ and we have confirmed that the real space exact diagonalization spectra using $N=441$ and thermodynamic limit results coincide. 
\section*{Integrals used within the harmonic approximation of Coulomb elements}
\begin{align}
    I_1=\int_0^{\infty}\frac{d\omega}{2\omega^2+1} \text{exp} \left[ -\frac{a_M}{\kappa^{1/2}}\left(\frac{\omega^2}{2\omega^2+1} \right)\right].
\end{align}
\begin{align}
    I_2=\int_0^{\infty}\frac{d\omega}{2\omega^2+1} \text{exp} \left[ -\frac{a_M}{4\kappa^{1/2}}\left(\frac{\omega^2}{2\omega^2+1} \right)\right].
\end{align}
\section*{Effect of the modulation potential on the phase boundary}
\begin{figure}[h!]
\centering
\includegraphics[width=0.8\linewidth]{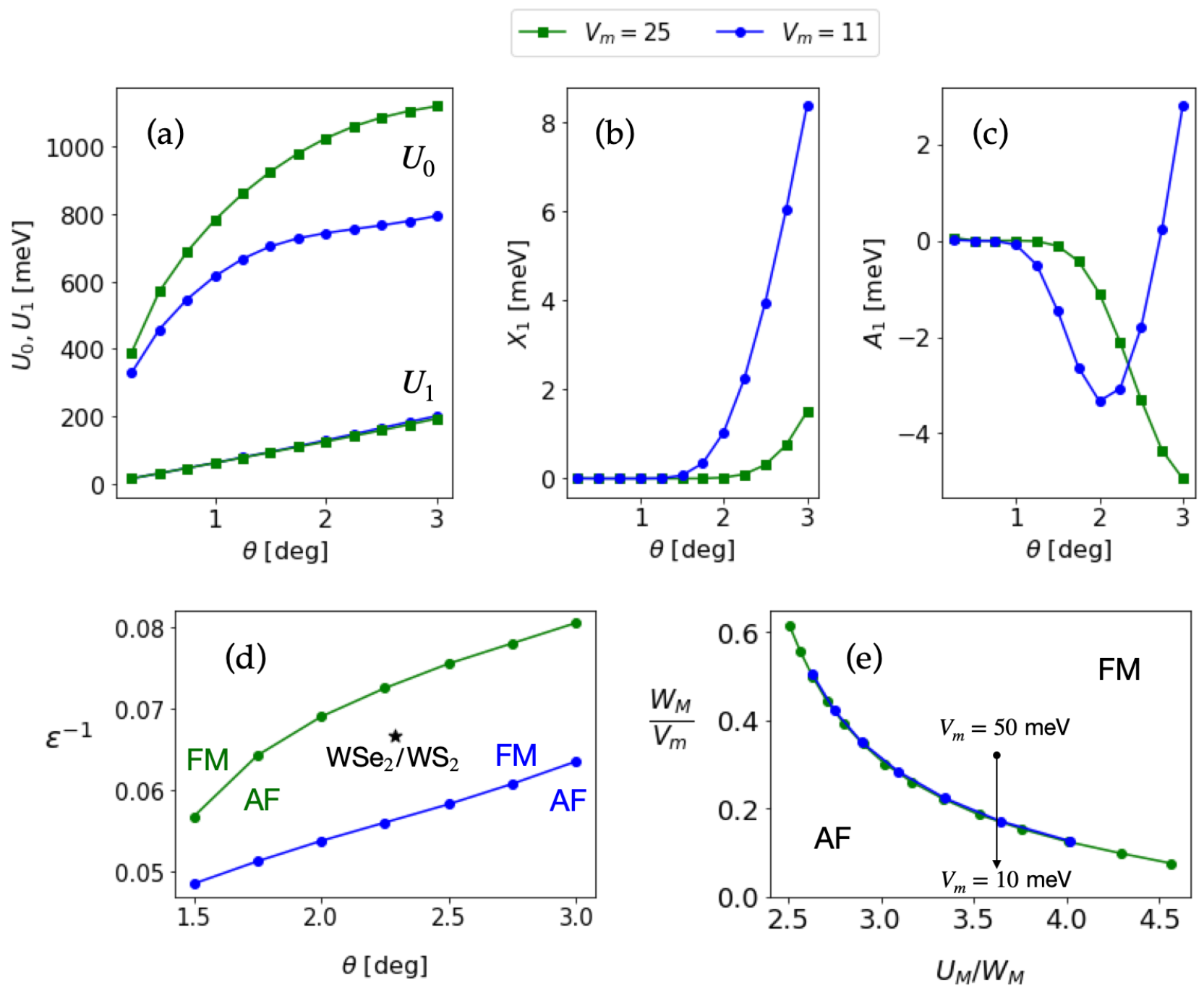}
\caption{Real space Coulomb elements as a function of twist angle, (a) onsite $U_0$ and nearest-neighbor direct interaction $U_1$, (b) exchange $X_1$, and (c) assisted hopping $A_1$, for $V_m=25$ meV and $V_m=11$ meV. We have fixed $\psi=-94^{\circ}$ and $\epsilon=1$. (d) 
Phase diagram of a $\nu=1$ twisted heterobilayer {\it vs.} interaction strength $\epsilon^{-1}$ and twist angle $\theta$, indicating the antiferromagnet-ferromagnet transition line for two values of $V_m$. The location in the phase diagram of a typical WSe$_2$/WS$_2$ sample with $\epsilon=15$ is indicated by a star. (e) Phase diagram in terms of dimensionless parameters indicating the universality of the  antiferromagnet to ferromagnet transition. The arrow indicates the trajectory of a WSe$_2$/WS$_2$ bilayer when $V_m$ is varied from 10 meV to 50 meV.}
\label{fig:Vm11_25}
\end{figure}

In the analysis presented in the main text, we focused on the dependence of the bilayer system on twist angle and dielectric constant variations. As mentioned previously, the geometry of the moir\'e superlattice (either honeycomb or triangular) depends on the particular material and determines the range of values that $\psi$ can take. Once this range is set, changes in the value of $\psi$ do not modify the physics of the bilayer significantly.  On the other hand, the effective value of $V_m$ can be modified by applying an external electric field or pressure to the sample, which will also modify the Wannier functions. In Fig. \ref{fig:Vm11_25}(a)-(c) we show the dependence of real space Coulomb elements on the twist angle for two values of the modulation potential strength, $V_m=25$ meV and $V_m=11$ meV (value used for calculations in the main text), for comparison. A larger value of the potential strength means that Wannier functions will be more localized, for that reason on-site interactions are stronger for $V_m=25$ meV but non-local interactions decrease in value. In Fig. \ref{fig:Vm11_25}(d) we show how changing the value of $V_m$ displaces the antiferromagnet-ferromagnet transition line in the phase diagram as a function of $\theta$ and $\epsilon^{-1}$. In order to relate our results to previous experiments we considered the case of aligned
WSe$_2$/WS$_2$, which has an effective twist angle $\theta=2.29^{\circ}$, as it is the most studied material so far. If we take the dielectric constant of the surrounding hBN to be $\epsilon\sim
5$ and include the effects of screening due to interband transitions and gates by making $\epsilon_{\text{eff}} \sim 15$, the location of this material in the phase diagram is indicated as a star in Fig. \ref{fig:Vm11_25}(d). It can be seen that by changing the modulation strength via an external field, one could tune between antiferromagnetism and ferromagnetism. 
The trends seen in Fig. \ref{fig:Vm11_25}(a)-(d) indicate that our conclusions apply for arbitrary values of the modulation potential, given it is strong enough to localize electrons. To illustrate this, we take the three relevant energy scales of the model: The kinetic energy scale $W_M=\hbar^2/m^*a_M^2$, the modulation strength $V_m$ and the interaction scale $U_M=e^2/\epsilon a_M$ and create a phase diagram in terms of two ratios between them, shown in Fig. \ref{fig:Vm11_25}(e). The antiferromagnet-ferromagnet transition lines for the two potential strengths coincide. This indicates that changing the modulation potential strength modifies the angle and value of interaction strength at which the transition happens, but that it is universal and should be present for any moir\'e TMD heterobilayer. The trajectory that an unrotated WSe$_2$/WS$_2$ sample would follow as the modulation strength is varied from 10 meV to 50 meV is indicated as an arrow in Fig. \ref{fig:Vm11_25}(e).

Additionally, from the color map in Fig. \ref{fig:AntiferroFerrotransition}(c) it can be seen that the energy scales associated to the effective spin models are larger for larger twist angles, which would facilitate the detection of magnetism. Assuming $\epsilon \sim 15$, for $\theta \sim 1^{\circ}$ we have $J_1 \sim 0.1$ mK, while for $\theta \sim 2.29^{\circ}$ $J_1 \sim 0.92$ K. The previous analysis suggests that tuning between an antiferromagnetic and a ferromagnetic Mott insulator is possible in semiconductor moir\'e materials. In order to measure this effect 1) larger angles are preferred, 2) the distance from the sample to metallic gates should be large in comparison to the moir\'e length in order to avoid further screening of the Coulomb interaction, 3) an applied electric field can be used to tune between the two phases. This sets the stage for future experimental confirmation of the phenomenon and also opens possibilities to detect spin liquids around the phase transition, as well as superconductivity and itinerant ferromagnets beyond half-filling. 
\section*{Couplings of the effective spin Hamiltonian}
We expand the real-space Hamiltonian in Eq. \eqref{ElectronInteractions} at half-filling to a low-energy effective spin Hamiltonian up to order $\tilde{t}_1^4/U_0^3$ \cite{AllanSW,AllanLargeU}. Our approximation of the effective spin model contains the first, second and third-neighbor hoppings $t_1, t_2, t_3$ and Coulomb interaction terms $U_0,U_1,U_2,U_3,X_1,X_2,X_3,A_1,A_2,A_3$ and $P_1$. We denote $\tilde{t}_n = t_n - A_n$ as the total $n$-th nearest neighbor hopping amplitude. The resulting Hamiltonian is given by Eq. \eqref{EffectiveSpin} in the main text  

\begin{align}
    H_{\text{eff}} &= \, J_1\sum_{\braket{i,j}}{\bf S}_i\cdot{\bf S}_j+J_2\sum_{\braket{\braket{i,j}}}{\bf S}_i\cdot{\bf S}_j
    +J_3\sum_{\braket{\braket{\braket{i,j}}}}{\bf S}_i\cdot{\bf S}_j \nonumber \\
    &+\sum_{\ring}J_4^1\left[({\bf S}_1\cdot{\bf S}_2)({\bf S}_3\cdot{\bf S}_4)+({\bf S}_1\cdot{\bf S}_4)({\bf S}_3\cdot{\bf S}_2)\right]
    +J_4^2({\bf S}_1\cdot{\bf S}_3)({\bf S}_2\cdot{\bf S}_4), \nonumber
\end{align}
where the spin model couplings are expressed using real space the Coulomb interaction terms
\begin{align}
    J_1= -2X_1&+\frac{4\tilde{t}_1^2}{U_0-U_1}-\frac{4(P_1-X_1)\tilde{t}_1^2}{(U_0-U_1)^2}
     +\frac{8\tilde{t}_1^4}{(U_0-U_1)^3} \left(\frac{ U_0-U_1}{2 U_0-3 U_1+U_2} \right.\nonumber\\
     & \quad \left.+\frac{4 (U_0-U_1)}{2 U_0-U_1-U_2} +\frac{3 (U_0-U_1)}{U_0-U_2}+\frac{2 (U_0-U_1)}{U_0-U_3}-11\right),
\end{align}
\begin{align}
    J_2=-2X_2+\frac{4\tilde{t}_2^2}{U_0-U_2}+\frac{8 \tilde{t}_1^4}{(U_0-U_1)^3}\left(1-\frac{U_0-U_1}{U_0-U_2}+\frac{U_0-U_1}{2 U_0-3 U_1+U_2}\right),
\end{align}
\begin{align}
    J_3=-2X_3+\frac{4\tilde{t}_3^2}{U_0-U_3}-\frac{4 \tilde{t}_1^4}{(U_0-U_1)^3}\left(\frac{U_0-U_1}{U_0-U_3}-2\right),
\end{align}
\begin{align}
    J_4^1=\frac{32 \tilde{t}_1^4}{(U_0-U_1)^3} \left(2+\frac{U_0-U_1}{U_0-U_2}-\frac{U_0-U_1}{2 U_0-U_1-U_2}\right),
\end{align}
\begin{align}
    J_4^2=-\frac{32\tilde{t}_1^4}{(U_0-U_1)^3}\left(1+\frac{U_0-U_1}{U_0-U_2}+\frac{U_0-U_1}{2 U_0-3 U_1+U_2}\right).
\end{align}
Notice that when $U_0\gg U_1,U_2,U_3$ and $X_1, A_1$ are not considered, we recover the usual expressions for the first-neighbor and ring-exchange Heisenberg couplings $J_1=4(t_1^2/U_0)\left[1-7(t_1^2/U_0)^2 \right]$ and $J_4^1=-J_4^2=80t_1^4/U_0^3$ \cite{AllanSW}.

\end{document}